\documentclass{epl}
\usepackage{amsmath}

\title{Quasiparticle transport in arrays of chaotic cavities}
\shorttitle{Transport in arrays of chaotic cavities}
\author{Mihajlo Vanevi\' c\inst{1} \and Wolfgang Belzig\inst{2}}

\institute{
  \inst{1} Departement Physik und Astronomie, Universit\" at Basel
           \\ Klingelbergstrasse 82, CH-4056 Basel, Switzerland\\
  \inst{2} Departement Physik, Universit\" at Konstanz
           \\ D-78457 Konstanz, Germany
}

\pacs{72.70.+m}{Noise processes and phenomena}
\pacs{73.23.-b}{Electronic transport in mesoscopic systems}
\pacs{74.50.+r}{Tunneling phenomena; point contacts, weak links, Josephson effects}

\newcommand{\vG}{{\check G}}
\newcommand{\vcG}{{\check{\cal G}}}
\newcommand{\vgg}{{\check \gamma}}
\newcommand{\vgt}{{\check \tau}}
\newcommand{\cT}{{\cal T}}

\newcommand{\gre}{{\cal E}}

\newcommand{\upi}{{\mathrm{i}}}

\newcommand{\Tr}{{\mathop{\rm tr}\nolimits}}
\renewcommand{\Im}{{\mathop{\rm Im}\nolimits}}
\newcommand{\arccosh}{{\mathop{\rm arccosh}\nolimits}}
\newcommand{\Com}[2]{{[ #1, #2 ]}}
\newcommand{\aCom}[2]{{\{ #1, #2 \}}}

\begin{document}

\maketitle

\begin{abstract}
We find the distribution of transmission eigenvalues in a series
of identical junctions between chaotic cavities using the circuit
theory of mesoscopic transport. This distribution rapidly
approaches the diffusive wire limit as the number of junctions
increases, independent of the specific scattering properties of a
single junction. The cumulant generating function and the first three
cumulants of the charge transfer through the system are obtained
both in the normal and in the superconducting state.
\end{abstract}

\section{Introduction}
Higher-order correlators of current fluctuations in mesoscopic
conductors have been studied extensively over the last decade
both theoretically
\cite{art:BuettikerPRB92,art:deJongBeenakker1997,art:BeenakkerRevModPhys,%
art:BlanterButtikerPhysRep00,art:BlanterCONDMAT05,nazarov:book,%
book:NanophysicsCohAndTransport05}
and experimentally
\cite{art:SchoenenbergerPRB02,art:SongPRL06,art:SchoelkopfPRL97,%
art:OberholzerPRB99,art:JehlNature00,art:ProberPRL00,art:HoffmannPRL03,%
art:Choi,art:CMarcusCONDMAT06,art:ProberPRL03,art:Bomze}. The reason
is that they contain, in general, additional information to the usual
differential conductance such as higher moments of the transmission
eigenvalue distribution, the value of effective charge involved in
transport processes, the size of internal energy scales of the system
or the correlations intrinsic to the many-body state of entangled
systems
\cite{art:MartinCONDMAT05,art:ReuletCONDMAT05,art:BeenakkerCONDMAT05}.
While the conductance is proportional to the average transmission
probability of the structure at low temperatures, the current noise
power $P_I$ depends on the second moment of transmission eigenvalue
distribution which is characterized by the Fano factor
$F=P_I/2eI=[\sum_n T_n(1-T_n)]/\sum_n T_n$. Here $e$ is the electron
charge, $I$ is the average current through the sample, and $T_n$ are the
transmission eigenvalues. Recent experiments on noise confirmed the
theoretical predictions
\cite{art:deJongBeenakker1997,art:BeenakkerRevModPhys} on the
universal distributions of transmission eigenvalues in a metallic
diffusive wire \cite{art:SchoelkopfPRL97,art:OberholzerPRB99} and in
an open chaotic cavity \cite{art:SchoenenbergerPRB02} with Fano
factors $F=1/3$ and $F=1/4$, respectively. The crossover from a single
cavity to the diffusive wire limit as the number of internal junctions
increases was studied experimentally by Oberholzer \etal
\cite{art:SchoenenbergerPRB02} and Song \etal \cite{art:SongPRL06}
recently.

Particle-hole correlations introduced by a superconducting terminal
also modify the noise. The low temperature noise of the subgap
transport is doubled for tunnel junctions \cite{art:HoffmannPRL03}
and in diffusive normal wires in contact with a superconductor
\cite{art:JehlNature00,art:ProberPRL00}. The noise in an open
cavity is found to be more than two times larger in the
superconducting state \cite{art:Choi} than in the corresponding
normal state junction, in agreement with theoretical predictions
\cite{art:deJongBeenakker1997}.

The third correlator contains the first three moments of transmission
eigenvalue distribution and is related to the asymmetry of the
distribution \cite{art:Bulashenko}. In contrast to the current noise
which is thermally dominated at temperatures larger than the bias
voltage according to the fluctuation-dissipation theorem, the third
correlator is in this regime proportional to the current, without the need
to correct for the thermal noise. However, higher-order correlators
are increasingly more difficult to measure because of the statistical
fluctuations \cite{art:LevitovReznikovPRB04} and the influence of
environment \cite{art:ProberPRL03,art:ReuletCONDMAT05}.  Recent
measurements of the third-order correlations of voltage fluctuations
across the nonsuperconducting tunnel junctions by Bomze {\etal}
\cite{art:Bomze} confirmed the Poisson statistics of electron transfer
at negligible coupling of the system to environment.

The statistical theory of transport, full counting statistics
\cite{art:LevitovJETP,art:LeeLevitovYakovetsPRB95}, provides the
most detailed description of charge transfer in mesoscopic
conductors. The semiclassical cascade approach to higher-order
cumulants based on the Boltzmann-Langevin equations has been developed
by Nagaev \etal \cite{art:Nagaev}. The stochastic path integral
theory of full counting statistics was introduced by Pilgram \etal
\cite{art:JordanSukhorukovPilgram}. The quantum-mechanical
theory of full counting statistics based on the extended
Keldysh-Green's function
technique \cite{art:nazarovANNPHYS99,art:BelzigPRL01,art:BelzigPRL01a}
in the discretized form of the circuit
theory \cite{art:Nazarov95,art:NazarovSuperlatt99}
was put forward for multiterminal circuits by Nazarov and
Bagrets\cite{art:NazarovBagretsPRL02}.
%
In this article we use the circuit theory
\cite{art:NazarovHTCN05,art:BelzigInNazarovBook} to study the
elastic quasiparticle transport in arrays of chaotic cavities
focusing on the crossover from a single cavity to the universal
limit of a diffusive wire \cite{art:NazarovPRL94} as the number of
inner contacts increases. We find the analytical expressions for
the distribution of transmission eigenvalues, the cumulant
generating function and the first three cumulants both in the
normal and in the superconducting state, generalizing the previous
results on noise \cite{art:SchoenenbergerPRB02} in such a system
to all higher-order correlations. The similar finite-size effects
on the noise and the third correlator have been studied
numerically by Roche and Dou\c cot \cite{art:RocheDoucotEPJB02}
within an exclusion model. Ballistic to diffusive crossover in
metallic conductors with obstacles as a function of increasing
disorder has been studied by Mac\^ edo \cite{art:MacedoPRB02}
within the scaling theory of transport combined with the circuit
theory. The effects of Coulomb interaction on the current and
noise in chaotic cavities and diffusive wires have been studied by
Golubev \etal \cite{art:GolubevZaikinPRB04}. The noise in series
of junctions has been measured by Oberholzer \etal
\cite{art:SchoenenbergerPRB02} and Song \etal \cite{art:SongPRL06}
recently.

\section{Transport in an array of chaotic cavities}

\begin{figure}[t]
\onefigure[height=2.7 cm]{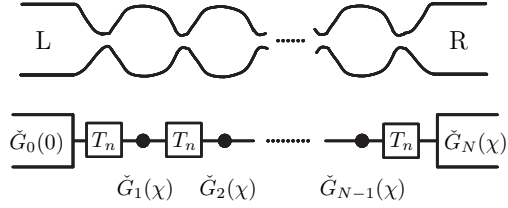} \caption{An array of
chaotic cavities in series between the Fermi leads (top) and the
discrete circuit-theory representation of the system (bottom). The
leads and the cavities are associated with the corresponding
matrix Greens functions $\vG_i$. The junctions are characterized
by a set of transmission eigenvalues $\{T_n\}$. The matrix current
is conserved throughout the circuit.} \label{fig:cavities}
\end{figure}

The system we consider consists of chaotic cavities in
series between $N$ identical junctions characterized by $N_{ch}$
transverse channels with transmission eigenvalues $\{ T_n \}$. We
can neglect the energy dependence of the transmission eigenvalues
of the system if the electron dwell time is small with respect to
time scales set by the inverse temperature and applied voltage.
Also we neglect the charging effects assuming that the
conductances of the junctions are much larger than the conductance
quantum $2e^2/h$. The quasiparticle distribution function is
isotropic between junctions due to chaotic scattering in the
cavities.
We apply the circuit theory of mesoscopic transport and represent the
specific parts of the system by the corresponding discrete circuit
elements, as shown in fig.~\ref{fig:cavities}. The Greens functions
of the leads are denoted as $\vG_0(0) \equiv \vG_L(0)$ and
$\vG_N(\chi) \equiv \vG_R(\chi)$, while the Greens functions of the
internal nodes are $\vG_i(\chi)$, $i=1,\ldots,N-1$. The counting field
$\chi$ can be incorporated through the boundary condition
\cite{art:BelzigInNazarovBook} at the right lead according to
\begin{equation}
\vG_N(\chi) = e^{-\upi(\chi/2)\vgt_K}\; \vG_N(0)\; e^{\upi(\chi/2)\vgt_K}.
\end{equation}
Here $\vG_{0,N}(0)$ are the bare Greens functions of the Fermi
leads in the Keldysh$(\bar{\;})$ $\otimes$ Nambu$(\hat{\;\;})$
space, $\bar\tau_i$ and $\hat\sigma_i$ are the Pauli matrices and
$\vgt_K = \bar\tau_1 \hat\sigma_3$. The connection between
adjacent nodes is described in general by a matrix current
\cite{art:NazarovSuperlatt99}
\begin{equation}
\check I_{i,i+1} = \frac{2e^2}{h}
\sum_{n=1}^{N_{ch}}
\frac{2T_n\Com{\vG_{i+1}}{\vG_i}}{4+T_n(\aCom{\vG_{i+1}}{\vG_i}-2)},
\end{equation}
which flows from the node $i$ to the node $i+1$. The set of
circuit-theory equations for the Greens functions of the internal
nodes consists of matrix current conservations $\check I_{i,i+1} =
const$ and normalization conditions $\vG_i^2=1$. As shown in ref.
\cite{art:Vanevic}, we can seek for the solution in the form
$\vG_i = \check F_{iL}(\aCom{\vG_0}{\vG_N})\;\vG_0 +\check
F_{iR}(\aCom{\vG_0}{\vG_N})\;\vG_N$, where the functions $\check
F_{iL,R}$ depend only on the anticommutator $\aCom{\vG_0}{\vG_N}$
and therefore commute with all $\vG_j$. As a consequence, the
anticommutators $\aCom{\vG_i}{\vG_j}$ depend only on
$\aCom{\vG_0}{\vG_N}$ and commute with all $\vG_k$. We emphasize
that the above consideration is independent of the concrete matrix
structure of the Greens functions and relies only on the
quasiclassical normalization conditions $\vG_i^2=1$. Because the
junctions are identical, the matrix current conservation reduces
to \cite{art:Vanevic}
\begin{equation}\label{eq:Gi}
\vG_i =
\frac{\vG_{i-1}+\vG_{i+1}}{\sqrt{\aCom{\vG_{i-1}}{\vG_{i+1}}+2}}.
\end{equation}
Taking the anticommutator of eq.~\eqref{eq:Gi} with $\vG_i$ and
$\vG_{i+1}$, respectively, we find that
$\aCom{\vG_{i-1}}{\vG_i}=\aCom{\vG_i}{\vG_{i+1}}\equiv \vcG'$, for
all $i$. Our aim is to find $\vcG'$ in terms of $\vcG \equiv
\aCom{\vG_0}{\vG_N}$. Now we take the anticommutator of eq.
\eqref{eq:Gi} with $\vG_0$ and obtain the following difference
equation
\begin{equation}\label{eq:difrcEq}
\vgg_{i+1}-\vcG'\vgg_i + \vgg_{i-1} = 0,
\end{equation}
where $\vgg_i=\aCom{\vG_0}{\vG_i}$. After solving eq.
\eqref{eq:difrcEq} with the boundary conditions $\vgg_0=2$ and
$\vgg_N=\vcG$, and using that $\vgg_1=\vcG'$, we find
\begin{equation}\label{eq:G'}
\vcG' = [(\vcG+\sqrt{\vcG^2-4})/2]^{1/N}
+
[(\vcG-\sqrt{\vcG^2-4})/2]^{1/N}.
\end{equation}
%
%

The cumulant generating function $S(\chi)$ of charge transfer through
the structure can be obtained as a sum of the actions
of the connected pairs of nodes \cite{art:NazarovBagretsPRL02}.
For identical junctions in series $S(\chi)$ reduces to the contribution
of a single junction multiplied by $N$:
\begin{equation}\label{eq:S}
S(\chi) = -\frac{t_0}{2h}\int \upd\gre \; \Tr[\tilde S(\chi)],
\qquad \text{with} \qquad
\tilde S(\chi) = N \sum_{n=1}^{N_{ch}} \ln [1+T_n(\vcG'-2)/4].
\end{equation}
Here $t_0$ is the total measurement time which is
much larger than the characteristic time scale on which the current
fluctuations are correlated. For a large number of
junctions, $N \gg 1$, the cumulant generating function given by
eq.~\eqref{eq:S} approaches the universal limit of a diffusive wire
\cite{art:LeeLevitovYakovetsPRB95,art:Nazarov95}
$
S(\chi)= (-t_0\tilde g/8h)
\int \upd \gre\; \Tr[\arccosh^2(\vcG/2)]
$,
which does not depend on the specific scattering properties $\{T_n\}$
of a single junction, the shape of the conductor or the impurity
distribution
\cite{art:NazarovPRL94,art:NazarovHTCN05}.
Here $\tilde g=(\sum_n T_n)/N$ is the total conductance of a wire
in units of $2e^2/h$.

The distribution of transmission eigenvalues of a composite junction
$\rho_{N,\{T_n\}}(\cT)$ is directly related to the cumulant generating
function by \cite{art:Nazarov95}
\begin{equation}\label{eq:rho}
\rho_{N,\{T_n\}}(\cT) = \frac{1}{\pi \cT^2} \; \Im
\Big (
\frac{\partial \tilde S(\zeta)}{\partial \zeta}
\Big \vert_{\zeta = - 1/\cT -\upi\, 0}
\Big ),
\end{equation}
where $\zeta=(\vcG-2)/4$. From eqs.~\eqref{eq:G'},
\eqref{eq:S} 
and \eqref{eq:rho} we find
\begin{multline}\label{eq:rho-junctions}
\rho_{N,\{T_n\}}(\cT)
=
\rho_D(\cT)\;
\frac{\sin(\pi/N)}{\pi/N}
\\
\times
\left\langle
\frac{4T_n[(b^{1/N}+b^{-1/N})(2-T_n)+2T_n \cos(\pi/N)]}
{[(b^{1/N}+b^{-1/N})(2-T_n)+2T_n \cos(\pi/N)]^2
-4(1-T_n)(b^{1/N}-b^{-1/N})^2}
\right\rangle.
\end{multline}
Here $\rho_D(\cT)=(\tilde g/2)(1/\cT\sqrt{1-\cT})$ is the
transmission distribution of a diffusive wire,
$b=(1+\sqrt{1-\cT})^2/\cT$, and $\langle\cdots \rangle=(\sum_n
\cdots)/(\sum_n T_n)$ denotes the averaging over the transmission
eigenvalues of a single junction. For $N$ open contacts in series
($T_n=1$) eq.~\eqref{eq:rho-junctions} reduces to
\begin{equation}\label{eq:rho-cavities}
\rho_{N}(\cT)
=
\rho_D(\cT)\;
\frac{\sin(\pi/N)}{\pi/N}
\frac{4 \cT^{1/N}}{(1+\sqrt{1-\cT})^{2/N}+(1-\sqrt{1-\cT})^{2/N}
+2\cT^{1/N}\cos (\pi/N)}.
\end{equation}%
%
\begin{figure}[t]
\onefigure[width=12cm, height=4cm]{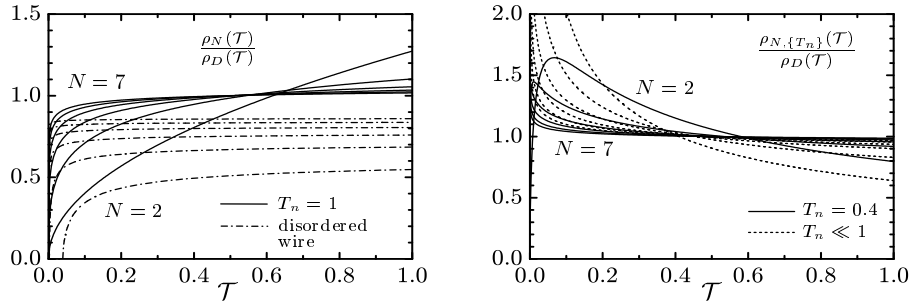}
\caption{Distribution of transmission eigenvalues for $N$ open
contacts in series (left panel) and for contacts of lower
transparency $T_n=0.4$ and $T_n\ll 1$ (right panel), normalized to
the transmission distribution of a diffusive wire $\rho_D(\cT)$.
Ballistic to diffusive crossover in a metallic disordered wire is
shown for comparison (left panel, dash-dotted curves) as a
function of increasing disorder $L/l = N-1$.} \label{fig:plots}
\end{figure}%
%
The crossover from a single cavity to the diffusive regime as the
number of junctions $N$ increases is shown in fig.~\ref{fig:plots}
for open contacts (left panel) and for contacts of lower
transparency $T_n=0.4$ and $T_n\ll 1$ (right panel). The
transmission distribution of a metallic disordered wire of the
length $L$ and the mean free path $l$ is shown for comparison
(dash-dotted curves in fig.~\ref{fig:plots}; cf.
ref.~\cite{art:BeenakkerRevModPhys}).

The transmission eigenvalue distributions given by
eqs. \eqref{eq:rho-junctions} and \eqref{eq:rho-cavities} can be probed
experimentally by measuring higher-order correlators of current
fluctuations across the junction at low temperatures. The first
three moments of charge transport statistics are related to the
average current, the current noise power and the third correlator
according to $I=\upi(e/t_0)\partial_\chi S|_{\chi=0}$, $P_I =
(2e^2/t_0)\partial^2_\chi S|_{\chi=0}$, and
$C_I=-\upi(e^3/t_0)\partial^3_\chi S|_{\chi=0}$, respectively.  In
the linear regime, which we consider here, the current is
proportional to the bias voltage with conductance given by $\tilde
g=(\sum_n T_n)/N=\int_0^1 \upd\cT \rho_{N,\{T_n\}}(\cT)\;\cT$ in
units of $2e^2/h$. At temperatures much lower than the bias
voltage, the current noise power and the third correlator are
linear in the current, with the slopes given by the Fano factor
$
F=\partial P_I/\partial(2eI) = (1/\tilde g) \int_0^1 \upd\cT
\rho_{N,\{T_n\}}(\cT) \cT(1-\cT)
$
and the "skewness"
$
C=\partial C_I/\partial(e^2I)= (1/\tilde g) \int_0^1 \upd\cT
\rho_{N,\{T_n\}}(\cT)\\ \times\cT(1-\cT)(1-2\cT) $, respectively.
For the normal-state junction the two parameters are given by
\vspace*{-1mm}
\begin{equation}\label{eq:FanoN}
F=\frac{1}{3}\left(1+\frac{2-3\langle T_n^2 \rangle}{N^2}\right)
\end{equation}
and 
\vspace*{-1mm}
\begin{equation}\label{eq:C}
C = \frac{1}{15}
\left(
1+\frac{5(2-3\langle T_n^2 \rangle)}{N^2}
+ \frac{4-30\langle T_n^2(1-T_n) \rangle}{N^4}
\right).
\end{equation}
The Fano factor given by eq.~\eqref{eq:FanoN} coincides with the
result previously obtained within Boltz\-mann-Langevin approach
which takes into account both cavity noise and partition noise at
the contacts and was confirmed experimentally for up to three open
contacts in series \cite{art:SchoenenbergerPRB02}. The sign of $C$
is related to the asymmetry of transmission distribution
\cite{art:Bulashenko}, being negative (positive) when the weight
of the distribution is shifted towards open (closed) transmission
channels. Equation \eqref{eq:C} shows that closed channels prevail
in the composite junction for $N>2$ even for completely open inner
contacts, in agreement with eq.~\eqref{eq:rho-cavities}.

\begin{figure}[t]
\onefigure[width=12cm, height=4cm]{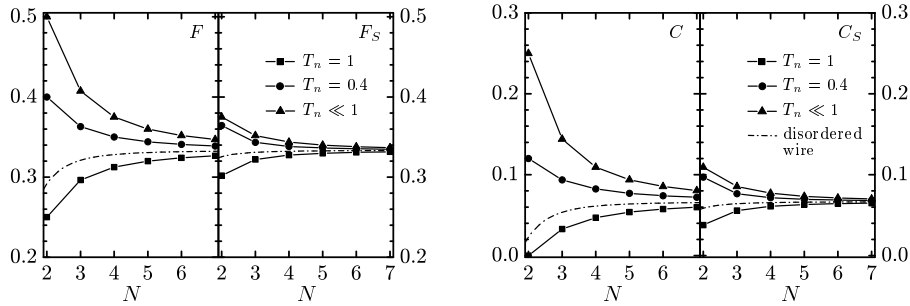} \caption{The
Fano factor $F$ (left panel) and the skewness $C$ (right panel) as
a function of the number of contacts in series $N$, shown for
different contact transparencies. The corresponding distributions
of transmission eigenvalues of the composite junctions are shown
in fig.~\ref{fig:plots}. The Fano factors $F_S$ and the skewnesses
$C_S$ (normalized by $e^*=2e$) of the superconducting junctions
are given for comparison. Ballistic to diffusive crossover in a
disordered wire ($L/l=N-1$) is shown by dash-dotted curves.}
\label{fig:FandC}
\end{figure}

Now we focus on the junction sandwiched between a normal-metal and
a superconductor in the coherent regime in which we can neglect
the particle-hole dephasing ($E_{th}\gg |eV|, k_BT, \Delta$ where
$E_{th}$ is the inverse dwell time). At temperatures and bias
voltages smaller than the superconducting gap $\Delta$, the
transport properties can be obtained by integrating the Andreev
reflection probability $R_A=\cT^2/(2-\cT)^2$ over the transmission
distribution \cite{art:deJongBeenakker1997,art:Vanevic} and
correcting for the effective charge $e^*=2e$. The conductance, the
Fano factor and the skewness are given by
$\tilde g_S=\int_0^1 \upd\cT\;\rho_{N,\{T_n\}}(\cT)\;R_A$
 (in units of $4e^2/h$),
$
F_S=\partial P_I/\partial(2e^*I) =
(1/\tilde g_S)
\int_0^1 \upd\cT \rho_{N,\{T_n\}}(\cT) R_A(1-R_A)
$,
and
$
C_S=\partial C_I/\partial(e^{*2}I)=
(1/\tilde g_S)
\int_0^1 \upd\cT \rho_{N,\{T_n\}}(\cT) R_A(1-R_A)(1-2R_A)
$, respectively.
For $N$ open contacts in series we find
\begin{equation}
\tilde g_S = \frac{N_{ch}}{2N} \frac{1}{\cos^2(\pi/4N)},
\qquad\quad
F_S = \frac{1}{3}
\Big[
1 - \frac{1}{4N^2}\Big(\frac{3}{\cos^2(\pi/4N)}-2 \Big)
\Big],
\end{equation}
and
\begin{equation}\label{eq:Cs}
C_S=\frac{1}{15}
\Big[
1 - \frac{5}{4N^2}\Big( \frac{3}{\cos^2(\pi/4N)}-2 \Big)
+ \frac{1}{8N^4} \Big( 2+15\frac{\sin^2(\pi/4N)}{\cos^4(\pi/4N)} \Big)
\Big].
\end{equation}
The skewness $C_S$ given by eq.~\eqref{eq:Cs} is positive for $N>1$
which indicates that the asymmetry of the Andreev reflection distribution
$\rho_{NA}(R_A)=\rho_N(\cT)d\cT/dR_A$ is in favor of closed
Andreev channels \cite{art:Vanevic}. For the more general case of
$N$ contacts characterized by transmission eigenvalues $\{T_n\}$
we obtain
$\tilde g_S = (\sum_n \alpha_n)/N$ and
$F_S = 1-(\sum_n\alpha_n\beta_n)/(\sum_n \alpha_n)$,
where
$\alpha_n=\sqrt{R^A_n}[\sqrt{R^A_n}+\cos(\pi/2N)]
/[1+\sqrt{R^A_n}\cos(\pi/2N)]^2$,
$\beta_n=2/3-(1/6N^2)\{1-6\alpha_n+
3\sqrt{R^A_n}/[\sqrt{R^A_n}+\cos(\pi/2N)]\}$, and $R^A_n=T_n^2/(2-T_n)^2$.
Fano factors and skewnesses for the normal-state and
superconducting junctions are shown in fig.~\ref{fig:FandC} as a
function of the number of contacts in series and for different
contact transparencies. It is interesting to note that in the
coherent superconducting regime, which we consider here, the
higher-order correlators satisfy the approximate scaling relations
$F_S(N)\approx F(2N)$ and $C_S(N)\approx C(2N)$ which are exact
for incoherent Andreev transport \cite{art:BelzigSamuelssonEPL03}.
For large $N$ this results in the full reentrance of
transport properties of a diffusive wire in contact with a
superconductor \cite{art:BelzigPRL01a} as a function of the
particle-hole coherence.

%

\section{Conclusion}

We have studied the transport properties of several chaotic cavities
in series using the circuit theory of mesoscopic transport.  We
obtained the analytical expression for the distribution of
transmission eigenvalues of the composite junction as a function of
the number of contacts and the scattering properties of a single
contact. This distribution generalizes the previous results on noise
in such a system \cite{art:SchoenenbergerPRB02} to all higher-order
cumulants. As an example we found the first three cumulants of the
charge transfer statistics both for the normal-state junction and in
the case when one lead is superconducting. The sign of the third
cumulant at high bias can be used to probe the asymmetry of the
transmission eigenvalue distribution: it is negative (positive) when
the weight of the distribution is more on open (closed) transport
channels.  As the number of contacts increases, all transport
properties approach the universal limit of a diffusive wire
\cite{art:NazarovPRL94}. While the crossover from a few
cavities to the diffusive-wire limit has already been studied
through the noise in the normal state
\cite{art:SchoenenbergerPRB02,art:SongPRL06}, experimental
investigations of either higher-order correlators or cavities in
contact with a superconductor are still to come.

\acknowledgments MV acknowledges Dimitrije Stepanenko and
Christoph Bruder for useful comments. This work has been supported
by the Swiss NSF and the NCCR "Nanoscience" (MV) and by the
Deutsche Forschungsgemeinschaft through SFB 513 and the
Landesstiftung Baden-W\" urttem\-b\-erg through the Research
Network "Functional Nanostructures" (WB).


\end{document}